\documentclass[%
 aip,
 amsmath,amssymb,
 reprint,%
]{revtex4-1}

\usepackage{graphicx}
\usepackage{dcolumn}
\usepackage{bm}

\usepackage[utf8]{inputenc}
\usepackage[T1]{fontenc}
\usepackage{mathptmx}
\usepackage{etoolbox}

\usepackage{hyperref}

\makeatletter
\def\@email#1#2{%
 \endgroup
 \patchcmd{\titleblock@produce}
  {\frontmatter@RRAPformat}
  {\frontmatter@RRAPformat{\produce@RRAP{*#1\href{mailto:#2}{#2}}}\frontmatter@RRAPformat}
  {}{}
}%
\makeatother
\begin{document}

\preprint{AIP/123-QED}

\title[Enhanced THz emission from Pt-Al alloys]{Enhanced THz emission from spintronic emitters with Pt-Al alloys}


\author{Felix Janus}
\affiliation{ 
New Materials Electronics Group, Department of Electrical Engineering and Information Technology, Technical University of Darmstadt, Merckstr. 25, 64283 Darmstadt, Germany
}%

\author{Nicolas Beermann}%
\affiliation{ 
Fakultät für Physik, Universität Bielefeld, Universitätsstraße 25, 33501 Bielefeld, Germany
}

\author{Jyoti Yadav}%
\affiliation{ 
New Materials Electronics Group, Department of Electrical Engineering and Information Technology, Technical University of Darmstadt, Merckstr. 25, 64283 Darmstadt, Germany
}%
\affiliation{ 
Spin Dynamics Lab, Department of Physics, Indian Institute of Technology Delhi, New Delhi, Delhi 110016, India
}%
\author{Reshma Rajeev Lekha}
\affiliation{ 
New Materials Electronics Group, Department of Electrical Engineering and Information Technology, Technical University of Darmstadt, Merckstr. 25, 64283 Darmstadt, Germany
}%

\author{Wentao Zhang}%
\author{Hassan A. Hafez}%
\author{Dmitry Turchinovich}
\affiliation{ 
Fakultät für Physik, Universität Bielefeld, Universitätsstraße 25, 33501 Bielefeld, Germany
}

\author{Markus Meinert}
 \email{markus.meinert@tu-darmstadt.de}
\affiliation{ 
New Materials Electronics Group, Department of Electrical Engineering and Information Technology, Technical University of Darmstadt, Merckstr. 25, 64283 Darmstadt, Germany
}


\date{\today}

\begin{abstract}
Platinum (Pt) is the element with the largest spin Hall conductivity and is known as the most efficient spin-to-charge conversion material in spintronic THz emitters. By alloying with aluminum (Al), its resistivity can be substantially increased, exceeding $100\,\mu\Omega$cm. While the spin Hall conductivity is reduced by alloying, the relative resistivity increase surpasses the reduction of spin Hall conductivity and thereby enhances the spin Hall angle. We make use of this mechanism to improve the commonly used Pt-based spintronic THz emitter and demonstrate that an increase of 67\% in the THz emission amplitude can be achieved between 20\% and 30\% Al in Pt. We show that the enhanced THz emission amplitude is driven by the enhanced multilayer impedance due to the larger resistivity.
\end{abstract}

\maketitle

Spintronic THz emitters convert an ultrashort laser pulse into a broadband THz pulse via the inverse spin Hall effect.\cite{Jungwirth2012, Hoffmann2013,Seifert2016, Papaioannou2021, Kampfrath2013, Seifert2017, Alekhin2017} Prototypically, they consist of a bilayer of nonmagnetic (NM) thin films and ferromagnetic (FM) thin films with thicknesses of a few nanometers. To maximize the THz output, the spin Hall angle $\theta_\mathrm{SH}$, the electronic mean free path $\lambda$, the ferromagnetic and nonmagnetic layer resistivities $\rho_\mathrm{FM}$ and $\rho_\mathrm{NM}$ and the film thicknesses of the layers $t_\mathrm{NM}$ and $t_\mathrm{FM}$, respectively, need to be carefully balanced. The spectral output density is modeled as a function of THz frequency $\omega/2\pi$ in the thin-film limit as\cite{Seifert2016}
\begin{equation}\label{eq:THz}
E_\mathrm{THz}(\omega) = {A\cdot B(\omega)\cdot \lambda}\cdot\tanh\left(\frac{t_\mathrm{HM}}{2\lambda}\right)\cdot\theta_\mathrm{SH}\cdot Z(\omega),
\end{equation}
where $A$ is the pump-light absorptance, while the factor $B$ captures the photon-to-spin-current conversion efficiency and the detector response function. The charge-current-to-electric-field conversion is described by the multilayer impedance\cite{Seifert2016}
\begin{equation}\label{eq:impedance}
Z(\omega) = \frac{Z_0}{n_1(\omega) + n_2(\omega) + Z_0 \int_0^d \mathrm{d}z \sigma_{xx}(z, \omega)}
\end{equation}
where $n_1(\omega)$ and $n_2(\omega)$ are the refractive indices of air and the substrate, respectively, $Z_0 = 377\,\Omega$, and $\sigma_{xx}(z, \omega)$ is the in-plane conductivity of the material at depth $z$. For simplicity, we take $\sigma_{xx}$ as constant across the film thickness, such that the integral turns into a sum of the reciprocal resistances per square. This approximation is valid if $\lambda \ll t_\mathrm{NM/FM}$ and is fulfilled for the samples dicussed here because of their high resistivity. For isotropic materials with mainly intrinsic spin Hall effect, we can write the spin Hall angle (SHA) as $\theta_\mathrm{SH} = \sigma_\mathrm{SH} / \sigma_{NM} = \sigma_\mathrm{SH} \rho_{NM}$ with the spin Hall conductivity (SHC) $\sigma_\mathrm{SH}$ and the resistivity $\rho$.\cite{Jungwirth2012, Hoffmann2013} Transport theory shows that $\lambda \rho = \kappa = \mathrm{const.}$ is a property of the Fermi surface.\cite{Gall2016} For NM thickness greater than four times the mean free path $\lambda$, the spin accumulation term becomes $\tanh\left(\frac{t_\mathrm{HM}}{2\lambda}\right) \approx 1$. Thus, we can rewrite Eq. \ref{eq:THz} as $E_\mathrm{THz}(\omega) = A\cdot B(\omega)\cdot\kappa\cdot \sigma_\mathrm{SH}\cdot Z(\omega)$. In the following, we drop the frequency dependence of $Z$ and take the DC values as an approximation. If both $\sigma_\mathrm{SH}$ and $\kappa$ are only affected slightly by alloying, the resistivity increase should enhance the THz emission signal due to a higher multilayer impedance $Z$. Similarly, a thinner NM film increases $Z$, but reducing $t_\mathrm{NM}$ below $4\lambda$ leads to incomplete spin-to-charge conversion. Thus, a higher resistivity also allows for thinner $t_\mathrm{NM}$ by reducing $\lambda$.

The element with the highest SHC and largest THz output is platinum (Pt).\cite{Tanaka2008, Go2024} However, it typically has low resistivity and a long mean-free path of several nanometers.\cite{Sagasta2016} Thus, the multilayer impedance of Pt-based stacks is rather low compared to high-resistivity materials like $\beta$-W or $\beta$-Ta. A previous attempt to enhance the resistivity of Pt by alloying with Au resulted in a 20\% increase of the THz emission, compared to pure Pt, with a maximum resistivity around $60\,\mu\Omega$cm at a thickness of 3\,nm.\cite{Meinert2020} As was recently pointed out by McHugh et al.,\cite{McHugh2024} alloying Pt with Al to create an fcc solid solution would greatly enhance the resistivity, while still maintaining a high SHC in the Pt-rich section of the phase diagram. The spin-orbit torque of a Pt-Al alloy system was studied by Lau et al.\cite{Lau2021} with a focus on the formation of L1$_2$-structured Pt$_3$Al. On SiO$_2$ substrates, the authors found an fcc solid solution with an enhanced resistivity and enhanced damping-like spin-orbit torque. The binary phase diagram shows multiple crystalline phases and a terminal solid solution with a solubility of about 5\% Al in Pt at room temperature (14\% at 1510$^\circ$C), whereas Pt is insoluble in Al up to the melting point.\cite{McAlister1986} In the Materials Project database,\cite{Jain2013} formation energies up to $-1.13$\,eV/atom are found for Pt-Al, which indicates a high structural stability of the compounds.

In this letter, we report on spintronic THz emitters with Pt-Al alloys as the NM layer. Thin films multilayer structures of fused silica / Pt$_{1-x}$Al$_x$ 5.5\,nm / CoFeB 2.5\,nm / Ta 0.5\,nm / TaO$_x$ 1\,nm were deposited by magnetron co-sputtering. Here, CoFeB denotes the composition Co$_{40}$Fe$_{40}$B$_{20}$. An additional stack without the heavy metal film was made to determine the shunt resistance in the other films. The substrates were held at room temperature, the deposition pressure was $2 \times 10^{-3}$\,mbar and the base pressure was below $5 \times 10^{-8}$\,mbar. The deposition rates of all films were typically around 0.1\,nm/s. The substrates were rotated at 30\,rpm to avoid the formation of uniaxial magnetic anisotropy. Growth rates were determined by x-ray reflectivity and structural characterization was performed with x-ray diffraction (XRD).

\begin{figure}[t!]
\includegraphics[width=8.6cm]{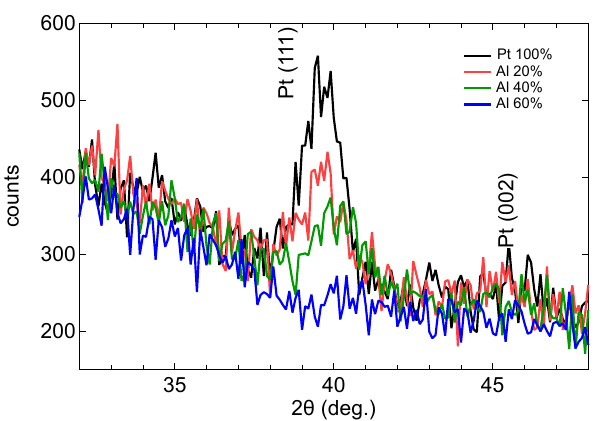}
\caption{\label{fig:XRD}
X-ray diffraction measurements with Cu K$_\alpha$ radiation and a double-bounce Ge monochromater. }
\end{figure}

Figure \ref{fig:XRD} shows the XRD measurements taken on selected THz emitter stacks. The reduction of the Pt (111) reflection originates from the reduction in the structure factor of the solid solution. While the films with up to 40\% Al are crystalline with an fcc structure and (111) texture, the sample with 60\% Al is amorphous. Applying a piecewise linear model for the background and Gaussians for the (111) reflection, we fit the diffraction data to extract the lattice constant and crystallite size. The Pt layer has a lattice constant of $a = (3.935 \pm 0.003)$\,\AA{}, slightly larger than the bulk value of 3.924\,\AA{}.\cite{McAlister1986}. The lattice constant (assuming an fcc lattice) decreases with increasing Al content to $(3.888 \pm 0.007)$\,\AA{} at 40\% Al. While this contradicts the Vegard rule (the lattice constant of Al is $4.05$\,\AA{}), it agrees with the phase diagram and previous observations.\cite{McAlister1986, Lau2021} This is an indication for the strong bonding between Pt and Al, which is much stronger as compared with typical intermetallic solid solutions to which the Vegard rule applies. The crystallite size is estimated as $(6.2 \pm 0.3)$\,nm for Pt and reduces to  $(5.0 \pm 0.5)$\,nm for 40\% Al. Both agree well with the nominal film thickness of 5.5\,nm. We note that microstrain is not considered here as an additional source of reflection broadening.

\begin{figure}[t!]
\includegraphics[width=8.6cm]{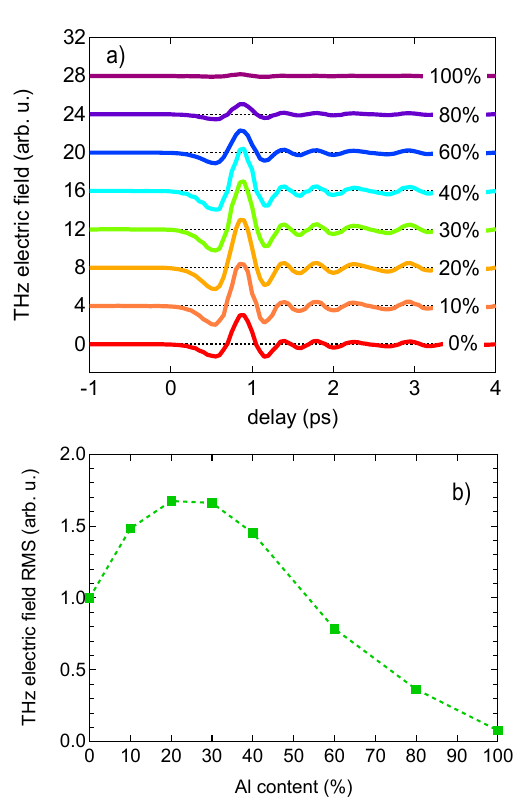}
\caption{\label{fig:THz_emission}
a) THz emission waveforms for all samples of the Pt-Al series. The waveforms are shifted vertically for clarity. The precentages represent the Al concentration. b) RMS averages of the waveforms, normalized to Pt.}
\end{figure}

\begin{figure}[t!]
\includegraphics[width=8.6cm]{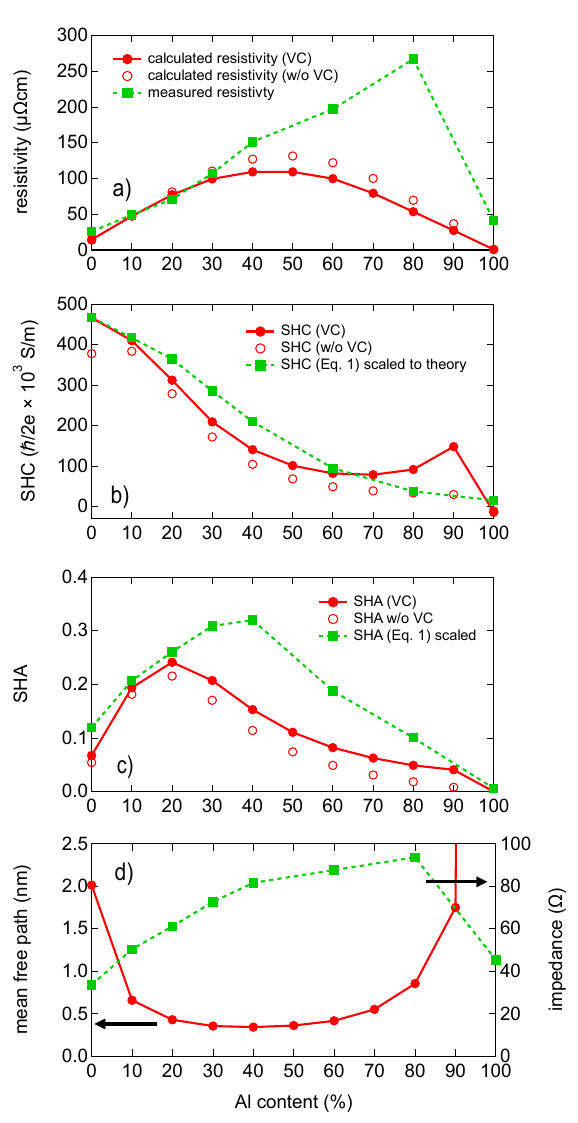}
\caption{\label{fig:exp_theory}
a) Measured and calculated resistivities of the Pt-Al alloy. Measured values use a parallel resistor model to correct for the shunting of the CFB layer. b) Calculated spin Hall conductivity; measured spin Hall conductivity extracted using Eq. \ref{eq:THz},  scaled to the calculated value with vertex corrections (VC). c) Calculated spin Hall angle as the product of calculated resistivity and SHC. Extracted spin Hall angle according to Eq. \ref{eq:THz}, scaled to $\sigma_{SH}(\mathrm{calc.}) \cdot \rho(\mathrm{expt.})$. d) Calculated mean free path (left axis) and multilayer impedance $Z$ according to the measured film resistivities (right axis).}
\end{figure}

For the THz emission measurements, we used an amplified Ti:sapphire laser system with a pulse width of 100\,fs centered at a wavelength of 800\,nm, a repetition rate of 1000\,Hz (chopped at 500\,Hz), and a power of 130\,mW. The samples were exposed to a magnetic field of 50\,mT to align the magnetization. The measurement of the THz electric field was performed with electro-optic sampling (EOS) in a 500$\mu$m ZnTe crystal and lock-in detection on a balanced detector. The laser illuminated the film through the substrate and the THz radiation was collected from the air-side with parabolic mirrors. The pump light absorptance was determined and found to be $A\approx 0.6 \pm 0.01$ for all samples. In Fig. \Ref{fig:THz_emission}, we show the THz waveforms (a) and the RMS average of the waveforms (b). We observe that the THz emission increases by up to 67\% with respect to pure Pt between 20\% and 30\% Al content. To apply the model equation \ref{eq:THz} and extract the spin Hall parameters, we measured the resistivity of the films. These were measured with a standard in-line four-point probing technique and a parallel resistor model was applied. Because the probe spacing is 3\,mm while the substrate size is $10 \times 10$\,mm$^2$ (with a film area of only $8 \times 8$\,mm$^2$), the usual analytical formula for the resistivity $\rho = (\pi / \mathrm{ln}2) R_{4w} t$, with the four-wire resistance R$_{4w}$ and the metal film thickness $t$ is not applicable here. We determined the prefactor of this formula by finite-element calculations and obtain $\rho = 2.17 R_{4w} t$ for central, diagonal placement of the four points across the square sample. The results are shown in Fig. \ref{fig:exp_theory} a). The measured resistivity increases from $26\,\mu\Omega$cm for the pure Pt film to approximately $260\,\mu\Omega$cm for 80\% Al. A high resistivity is in line with an amorphous structure for the Al-rich alloys. To better understand the resistivity of the fcc solid solution, we compare our results with density functional theory + Kubo-Bastin calculations with the SPR-KKR program.\cite{Ebert2011, Lowitzer2011, Ebert2015} Computational details are the same as in our previous publications.\cite{Meinert2020, Fritz2018} All calculations were performed with the alloy-analogy model for phonon contributions at 300\,K. We show the calculated quantities with and without vertex corrections (which cause the skew- and (partially) side-jump contributions).\cite{Chadova2015} Here we see that our thin-film resistivity of Pt is significantly higher than the calculated value ($14.4\,\mu\Omega$cm) and also larger than the experimental bulk value ($10.6\,\mu\Omega$cm). The main reasons for this are small crystallites and interface scattering (Fuchs-Sondheimer model) when the mean free path and the film thickness are comparable.\cite{Dutta2017} With higher Al content (up to 30\%), the experimental data and the calculations agree very well, with the resistivity of Pt$_{70}$Al$_{30}$ around $100\,\mu\Omega$cm. The theoretical curve complies with the typical Nordheim rule with higher Al content, whereas the experimental data deviate to the upside, indicating amorphous film growth. The pure Al film is highly resistive, which can be explained by the typical island growth of ultrathin Al films on glass substrates and the associated roughness.

To apply Eq. \ref{eq:THz}, we need a model for the mean free path. We use an interpolation between the $\kappa$ values of Pt and Al,\cite{Dutta2017, Gall2016} $\kappa(x) =  x \cdot 5.0 \times 10^{-16}~\Omega~\mathrm{m}^2 + (1-x) \cdot 2.9\times 10^{-16}~\Omega~\mathrm{m}^2$. The extracted SHC and SHA in comparison to the calculated data are shown in Fig. \ref{fig:exp_theory} b) and c). The measured spin Hall conductivity is overall in fair agreement with the calculated data.
Yet, initially, the calculated values decay faster and deviate from the trendline due to the VC on the Al-rich side.
In the experiment, such a divergence is not observed, which indicates the absence of an fcc solid solution.
This would agree with the insolubility of Pt in Al, whereas Al forms an fcc solid solution in Pt.\cite{McAlister1986} The SHA extracted from the THz data is scaled to the product of the calculated spin Hall conductivity and the experimental film resistivity. It extends significantly higher than the theoretical SHA (as the product of the calculated SHC and calculated resistivity) due to the higher film resistivity, which deviates from the Nordheim parabola. This result indicates that the SHA in our Pt$_{60}$Al$_{40}$ alloy could reach beyond 0.3. In Fig. \ref{fig:exp_theory} d) we show the theoretical electron mean free path according to our interpolation formula and the calculated resistivity. In addition, we also display the multilayer impedance $Z$ of our films. The mean free path is substantially decreased from 2\,nm in pure Pt to 0.34\,nm in Pt$_{60}$Al$_{40}$. It is thus much smaller than the film thickness in our experiment. This would enable much thinner emitter structures, where the Pt-Al layer may be as thin as 1.3\,nm while losing only 5\% of the full spin-to-charge-current conversion.

The driver of the enhanced THz emission $E_\mathrm{THz} \propto \kappa \cdot \sigma_\mathrm{SH}\cdot Z$ is the enhanced multilayer impedance $Z$, because $\kappa$ is only weakly increasing, while $\sigma_\mathrm{SH}$ is decaying more rapidly. Thus, the increase of the impedance remains as the main cause for the enhanced emission. We conclude that the sputter-deposited Pt-Al alloy system in the range up to $40\%$ Al forms an fcc solid solution and offers a high spin Hall conductivity, highly tunable resistivity, and short electron mean free path. It extends the possibilities for the design of spin Hall devices, where the pronounced shunting of the highly conductive Pt is an issue. At the same time, it allows for even thinner film structures with much higher spin-orbit torque. However, phase transformations into any of the many Pt-Al compounds, most importantly Pt$_3$Al, under process heat or current load may induce massive changes to the electrical properties of the films. However, the work on Pt$_3$Al by Lau et al.\cite{Lau2021} shows that this material can also be highly resistive and exhibits high SHC, such that a transformation into the L1$_2$ phase may not be detrimental to the transport properties. Our work demonstrates that a Pt-Al alloy can be used as an improved drop-in replacement for Pt in spintronic devices.

\begin{acknowledgments}
This work was supported by the Deutsche Forschungsgemeinschaft (DFG) under Project Numbers 513154775, 518575758, and under the Major Instrumentation Programme Project Numbers 468939474, and 511340083.

Bielefeld group acknowledges the financial support from the European Union’s Horizon 2020 research and
innovation program (Grant Agreement No. 964735 EXTREME-IR), Deutsche Forschungsgemeinschaft (DFG) within
Project No. 468501411-SPP2314 INTEGRATECH and Project No. 518575758 HIGHSPINTERA, Bundesministerium für Bildung und Forschung (BMBF) within Project No. 05K2022 PBA Tera-EXPOSE, and Bielefelder Nachwuchsfond.

\end{acknowledgments}

\section*{AUTHOR DECLARATIONS }

\subsection*{Conflict of Interest}
The authors have no conflicts to disclose.










\section*{Data Availability Statement}

The data that support the findings of this study are available upon reasonable request from the corresponding author.

\section*{References}

\bibliography{cite} 

\begin{thebibliography}{22}%
\makeatletter
\providecommand \@ifxundefined [1]{%
 \@ifx{#1\undefined}
}%
\providecommand \@ifnum [1]{%
 \ifnum #1\expandafter \@firstoftwo
 \else \expandafter \@secondoftwo
 \fi
}%
\providecommand \@ifx [1]{%
 \ifx #1\expandafter \@firstoftwo
 \else \expandafter \@secondoftwo
 \fi
}%
\providecommand \natexlab [1]{#1}%
\providecommand \enquote  [1]{``#1''}%
\providecommand \bibnamefont  [1]{#1}%
\providecommand \bibfnamefont [1]{#1}%
\providecommand \citenamefont [1]{#1}%
\providecommand \href@noop [0]{\@secondoftwo}%
\providecommand \href [0]{\begingroup \@sanitize@url \@href}%
\providecommand \@href[1]{\@@startlink{#1}\@@href}%
\providecommand \@@href[1]{\endgroup#1\@@endlink}%
\providecommand \@sanitize@url [0]{\catcode `\\12\catcode `\$12\catcode
  `\&12\catcode `\#12\catcode `\^12\catcode `\_12\catcode `\%12\relax}%
\providecommand \@@startlink[1]{}%
\providecommand \@@endlink[0]{}%
\providecommand \url  [0]{\begingroup\@sanitize@url \@url }%
\providecommand \@url [1]{\endgroup\@href {#1}{\urlprefix }}%
\providecommand \urlprefix  [0]{URL }%
\providecommand \Eprint [0]{\href }%
\providecommand \doibase [0]{http://dx.doi.org/}%
\providecommand \selectlanguage [0]{\@gobble}%
\providecommand \bibinfo  [0]{\@secondoftwo}%
\providecommand \bibfield  [0]{\@secondoftwo}%
\providecommand \translation [1]{[#1]}%
\providecommand \BibitemOpen [0]{}%
\providecommand \bibitemStop [0]{}%
\providecommand \bibitemNoStop [0]{.\EOS\space}%
\providecommand \EOS [0]{\spacefactor3000\relax}%
\providecommand \BibitemShut  [1]{\csname bibitem#1\endcsname}%
\let\auto@bib@innerbib\@empty
\bibitem [{\citenamefont {Jungwirth}, \citenamefont {Wunderlich},\ and\
  \citenamefont {Olejník}(2012)}]{Jungwirth2012}%
  \BibitemOpen
  \bibfield  {author} {\bibinfo {author} {\bibfnamefont {T.}~\bibnamefont
  {Jungwirth}}, \bibinfo {author} {\bibfnamefont {J.}~\bibnamefont
  {Wunderlich}}, \ and\ \bibinfo {author} {\bibfnamefont {K.}~\bibnamefont
  {Olejník}},\ }\bibfield  {title} {\enquote {\bibinfo {title} {Spin hall
  effect devices},}\ }\href {\doibase 10.1038/nmat3279} {\bibfield  {journal}
  {\bibinfo  {journal} {Nature Materials}\ }\textbf {\bibinfo {volume} {11}},\
  \bibinfo {pages} {382--390} (\bibinfo {year} {2012})}\BibitemShut {NoStop}%
\bibitem [{\citenamefont {Hoffmann}(2013)}]{Hoffmann2013}%
  \BibitemOpen
  \bibfield  {author} {\bibinfo {author} {\bibfnamefont {A.}~\bibnamefont
  {Hoffmann}},\ }\bibfield  {title} {\enquote {\bibinfo {title} {Spin hall
  effects in metals},}\ }\href {\doibase 10.1109/TMAG.2013.2262947} {\bibfield
  {journal} {\bibinfo  {journal} {IEEE Transactions on Magnetics}\ }\textbf
  {\bibinfo {volume} {49}},\ \bibinfo {pages} {5172--5193} (\bibinfo {year}
  {2013})}\BibitemShut {NoStop}%
\bibitem [{\citenamefont {Seifert}\ \emph {et~al.}(2016)\citenamefont
  {Seifert}, \citenamefont {Jaiswal}, \citenamefont {Martens}, \citenamefont
  {Hannegan}, \citenamefont {Braun}, \citenamefont {Maldonado}, \citenamefont
  {Freimuth}, \citenamefont {Kronenberg}, \citenamefont {Henrizi},
  \citenamefont {Radu}, \citenamefont {Beaurepaire}, \citenamefont {Mokrousov},
  \citenamefont {Oppeneer}, \citenamefont {Jourdan}, \citenamefont {Jakob},
  \citenamefont {Turchinovich}, \citenamefont {Hayden}, \citenamefont {Wolf},
  \citenamefont {Münzenberg}, \citenamefont {Kläui},\ and\ \citenamefont
  {Kampfrath}}]{Seifert2016}%
  \BibitemOpen
  \bibfield  {author} {\bibinfo {author} {\bibfnamefont {T.}~\bibnamefont
  {Seifert}}, \bibinfo {author} {\bibfnamefont {S.}~\bibnamefont {Jaiswal}},
  \bibinfo {author} {\bibfnamefont {U.}~\bibnamefont {Martens}}, \bibinfo
  {author} {\bibfnamefont {J.}~\bibnamefont {Hannegan}}, \bibinfo {author}
  {\bibfnamefont {L.}~\bibnamefont {Braun}}, \bibinfo {author} {\bibfnamefont
  {P.}~\bibnamefont {Maldonado}}, \bibinfo {author} {\bibfnamefont
  {F.}~\bibnamefont {Freimuth}}, \bibinfo {author} {\bibfnamefont
  {A.}~\bibnamefont {Kronenberg}}, \bibinfo {author} {\bibfnamefont
  {J.}~\bibnamefont {Henrizi}}, \bibinfo {author} {\bibfnamefont
  {I.}~\bibnamefont {Radu}}, \bibinfo {author} {\bibfnamefont {E.}~\bibnamefont
  {Beaurepaire}}, \bibinfo {author} {\bibfnamefont {Y.}~\bibnamefont
  {Mokrousov}}, \bibinfo {author} {\bibfnamefont {P.~M.}\ \bibnamefont
  {Oppeneer}}, \bibinfo {author} {\bibfnamefont {M.}~\bibnamefont {Jourdan}},
  \bibinfo {author} {\bibfnamefont {G.}~\bibnamefont {Jakob}}, \bibinfo
  {author} {\bibfnamefont {D.}~\bibnamefont {Turchinovich}}, \bibinfo {author}
  {\bibfnamefont {L.~M.}\ \bibnamefont {Hayden}}, \bibinfo {author}
  {\bibfnamefont {M.}~\bibnamefont {Wolf}}, \bibinfo {author} {\bibfnamefont
  {M.}~\bibnamefont {Münzenberg}}, \bibinfo {author} {\bibfnamefont
  {M.}~\bibnamefont {Kläui}}, \ and\ \bibinfo {author} {\bibfnamefont
  {T.}~\bibnamefont {Kampfrath}},\ }\bibfield  {title} {\enquote {\bibinfo
  {title} {Efficient metallic spintronic emitters of ultrabroadband terahertz
  radiation},}\ }\href {\doibase 10.1038/nphoton.2016.91} {\bibfield  {journal}
  {\bibinfo  {journal} {Nature Photonics}\ }\textbf {\bibinfo {volume} {10}},\
  \bibinfo {pages} {483--488} (\bibinfo {year} {2016})}\BibitemShut {NoStop}%
\bibitem [{\citenamefont {Papaioannou}\ and\ \citenamefont
  {Beigang}(2021)}]{Papaioannou2021}%
  \BibitemOpen
  \bibfield  {author} {\bibinfo {author} {\bibfnamefont {E.~T.}\ \bibnamefont
  {Papaioannou}}\ and\ \bibinfo {author} {\bibfnamefont {R.}~\bibnamefont
  {Beigang}},\ }\bibfield  {title} {\enquote {\bibinfo {title} {Thz spintronic
  emitters: A review on achievements and future challenges},}\ }\href {\doibase
  10.1515/nanoph-2020-0563} {\bibfield  {journal} {\bibinfo  {journal}
  {Nanophotonics}\ }\textbf {\bibinfo {volume} {10}},\ \bibinfo {pages}
  {1243--1257} (\bibinfo {year} {2021})}\BibitemShut {NoStop}%
\bibitem [{\citenamefont {Kampfrath}\ \emph {et~al.}(2013)\citenamefont
  {Kampfrath}, \citenamefont {Battiato}, \citenamefont {Maldonado},
  \citenamefont {Eilers}, \citenamefont {Nötzold}, \citenamefont {Mährlein},
  \citenamefont {Zbarsky}, \citenamefont {Freimuth}, \citenamefont {Mokrousov},
  \citenamefont {Blügel}, \citenamefont {Wolf}, \citenamefont {Radu},
  \citenamefont {Oppeneer},\ and\ \citenamefont {Münzenberg}}]{Kampfrath2013}%
  \BibitemOpen
  \bibfield  {author} {\bibinfo {author} {\bibfnamefont {T.}~\bibnamefont
  {Kampfrath}}, \bibinfo {author} {\bibfnamefont {M.}~\bibnamefont {Battiato}},
  \bibinfo {author} {\bibfnamefont {P.}~\bibnamefont {Maldonado}}, \bibinfo
  {author} {\bibfnamefont {G.}~\bibnamefont {Eilers}}, \bibinfo {author}
  {\bibfnamefont {J.}~\bibnamefont {Nötzold}}, \bibinfo {author}
  {\bibfnamefont {S.}~\bibnamefont {Mährlein}}, \bibinfo {author}
  {\bibfnamefont {V.}~\bibnamefont {Zbarsky}}, \bibinfo {author} {\bibfnamefont
  {F.}~\bibnamefont {Freimuth}}, \bibinfo {author} {\bibfnamefont
  {Y.}~\bibnamefont {Mokrousov}}, \bibinfo {author} {\bibfnamefont
  {S.}~\bibnamefont {Blügel}}, \bibinfo {author} {\bibfnamefont
  {M.}~\bibnamefont {Wolf}}, \bibinfo {author} {\bibfnamefont {I.}~\bibnamefont
  {Radu}}, \bibinfo {author} {\bibfnamefont {P.~M.}\ \bibnamefont {Oppeneer}},
  \ and\ \bibinfo {author} {\bibfnamefont {M.}~\bibnamefont {Münzenberg}},\
  }\bibfield  {title} {\enquote {\bibinfo {title} {Terahertz spin current
  pulses controlled by magnetic heterostructures},}\ }\href {\doibase
  10.1038/nnano.2013.43} {\bibfield  {journal} {\bibinfo  {journal} {Nature
  Nanotechnology}\ }\textbf {\bibinfo {volume} {8}},\ \bibinfo {pages}
  {256--260} (\bibinfo {year} {2013})}\BibitemShut {NoStop}%
\bibitem [{\citenamefont {Seifert}\ \emph {et~al.}(2017)\citenamefont
  {Seifert}, \citenamefont {Martens}, \citenamefont {Günther}, \citenamefont
  {Schoen}, \citenamefont {Radu}, \citenamefont {Chen}, \citenamefont {Lucas},
  \citenamefont {Ramos}, \citenamefont {Aguirre}, \citenamefont {Algarabel},
  \citenamefont {Anadón}, \citenamefont {Körner}, \citenamefont {Walowski},
  \citenamefont {Back}, \citenamefont {Ibarra}, \citenamefont {Morellón},
  \citenamefont {Saitoh}, \citenamefont {Wolf}, \citenamefont {Song},
  \citenamefont {Uchida}, \citenamefont {Münzenberg}, \citenamefont {Radu},\
  and\ \citenamefont {Kampfrath}}]{Seifert2017}%
  \BibitemOpen
  \bibfield  {author} {\bibinfo {author} {\bibfnamefont {T.}~\bibnamefont
  {Seifert}}, \bibinfo {author} {\bibfnamefont {U.}~\bibnamefont {Martens}},
  \bibinfo {author} {\bibfnamefont {S.}~\bibnamefont {Günther}}, \bibinfo
  {author} {\bibfnamefont {M.~A.~W.}\ \bibnamefont {Schoen}}, \bibinfo {author}
  {\bibfnamefont {F.}~\bibnamefont {Radu}}, \bibinfo {author} {\bibfnamefont
  {X.~Z.}\ \bibnamefont {Chen}}, \bibinfo {author} {\bibfnamefont
  {I.}~\bibnamefont {Lucas}}, \bibinfo {author} {\bibfnamefont
  {R.}~\bibnamefont {Ramos}}, \bibinfo {author} {\bibfnamefont {M.~H.}\
  \bibnamefont {Aguirre}}, \bibinfo {author} {\bibfnamefont {P.~A.}\
  \bibnamefont {Algarabel}}, \bibinfo {author} {\bibfnamefont {A.}~\bibnamefont
  {Anadón}}, \bibinfo {author} {\bibfnamefont {H.~S.}\ \bibnamefont
  {Körner}}, \bibinfo {author} {\bibfnamefont {J.}~\bibnamefont {Walowski}},
  \bibinfo {author} {\bibfnamefont {C.}~\bibnamefont {Back}}, \bibinfo {author}
  {\bibfnamefont {M.~R.}\ \bibnamefont {Ibarra}}, \bibinfo {author}
  {\bibfnamefont {L.}~\bibnamefont {Morellón}}, \bibinfo {author}
  {\bibfnamefont {E.}~\bibnamefont {Saitoh}}, \bibinfo {author} {\bibfnamefont
  {M.}~\bibnamefont {Wolf}}, \bibinfo {author} {\bibfnamefont {C.}~\bibnamefont
  {Song}}, \bibinfo {author} {\bibfnamefont {K.}~\bibnamefont {Uchida}},
  \bibinfo {author} {\bibfnamefont {M.}~\bibnamefont {Münzenberg}}, \bibinfo
  {author} {\bibfnamefont {I.}~\bibnamefont {Radu}}, \ and\ \bibinfo {author}
  {\bibfnamefont {T.}~\bibnamefont {Kampfrath}},\ }\bibfield  {title} {\enquote
  {\bibinfo {title} {Terahertz spin currents and inverse spin hall effect in
  thin-film heterostructures containing complex magnetic compounds},}\ }\href
  {\doibase 10.1142/S2010324717400100} {\bibfield  {journal} {\bibinfo
  {journal} {SPIN}\ }\textbf {\bibinfo {volume} {07}},\ \bibinfo {pages}
  {1740010} (\bibinfo {year} {2017})}\BibitemShut {NoStop}%
\bibitem [{\citenamefont {Alekhin}\ \emph {et~al.}(2017)\citenamefont
  {Alekhin}, \citenamefont {Razdolski}, \citenamefont {Ilin}, \citenamefont
  {Meyburg}, \citenamefont {Diesing}, \citenamefont {Roddatis}, \citenamefont
  {Rungger}, \citenamefont {Stamenova}, \citenamefont {Sanvito}, \citenamefont
  {Bovensiepen},\ and\ \citenamefont {Melnikov}}]{Alekhin2017}%
  \BibitemOpen
  \bibfield  {author} {\bibinfo {author} {\bibfnamefont {A.}~\bibnamefont
  {Alekhin}}, \bibinfo {author} {\bibfnamefont {I.}~\bibnamefont {Razdolski}},
  \bibinfo {author} {\bibfnamefont {N.}~\bibnamefont {Ilin}}, \bibinfo {author}
  {\bibfnamefont {J.~P.}\ \bibnamefont {Meyburg}}, \bibinfo {author}
  {\bibfnamefont {D.}~\bibnamefont {Diesing}}, \bibinfo {author} {\bibfnamefont
  {V.}~\bibnamefont {Roddatis}}, \bibinfo {author} {\bibfnamefont
  {I.}~\bibnamefont {Rungger}}, \bibinfo {author} {\bibfnamefont
  {M.}~\bibnamefont {Stamenova}}, \bibinfo {author} {\bibfnamefont
  {S.}~\bibnamefont {Sanvito}}, \bibinfo {author} {\bibfnamefont
  {U.}~\bibnamefont {Bovensiepen}}, \ and\ \bibinfo {author} {\bibfnamefont
  {A.}~\bibnamefont {Melnikov}},\ }\bibfield  {title} {\enquote {\bibinfo
  {title} {Femtosecond spin current pulses generated by the nonthermal
  spin-dependent seebeck effect and interacting with ferromagnets in spin
  valves},}\ }\href {\doibase 10.1103/PhysRevLett.119.017202} {\bibfield
  {journal} {\bibinfo  {journal} {Physical Review Letters}\ }\textbf {\bibinfo
  {volume} {119}},\ \bibinfo {pages} {017202} (\bibinfo {year}
  {2017})}\BibitemShut {NoStop}%
\bibitem [{\citenamefont {Gall}(2016)}]{Gall2016}%
  \BibitemOpen
  \bibfield  {author} {\bibinfo {author} {\bibfnamefont {D.}~\bibnamefont
  {Gall}},\ }\bibfield  {title} {\enquote {\bibinfo {title} {Electron mean free
  path in elemental metals},}\ }\href {\doibase 10.1063/1.4942216} {\bibfield
  {journal} {\bibinfo  {journal} {Journal of Applied Physics}\ }\textbf
  {\bibinfo {volume} {119}} (\bibinfo {year} {2016}),\
  10.1063/1.4942216}\BibitemShut {NoStop}%
\bibitem [{\citenamefont {Tanaka}\ \emph {et~al.}(2008)\citenamefont {Tanaka},
  \citenamefont {Kontani}, \citenamefont {Naito}, \citenamefont {Naito},
  \citenamefont {Hirashima}, \citenamefont {Yamada},\ and\ \citenamefont
  {Inoue}}]{Tanaka2008}%
  \BibitemOpen
  \bibfield  {author} {\bibinfo {author} {\bibfnamefont {T.}~\bibnamefont
  {Tanaka}}, \bibinfo {author} {\bibfnamefont {H.}~\bibnamefont {Kontani}},
  \bibinfo {author} {\bibfnamefont {M.}~\bibnamefont {Naito}}, \bibinfo
  {author} {\bibfnamefont {T.}~\bibnamefont {Naito}}, \bibinfo {author}
  {\bibfnamefont {D.~S.}\ \bibnamefont {Hirashima}}, \bibinfo {author}
  {\bibfnamefont {K.}~\bibnamefont {Yamada}}, \ and\ \bibinfo {author}
  {\bibfnamefont {J.}~\bibnamefont {Inoue}},\ }\bibfield  {title} {\enquote
  {\bibinfo {title} {Intrinsic spin hall effect and orbital hall effect in 4d
  and 5d transition metals},}\ }\href {\doibase 10.1103/PhysRevB.77.165117}
  {\bibfield  {journal} {\bibinfo  {journal} {Physical Review B}\ }\textbf
  {\bibinfo {volume} {77}},\ \bibinfo {pages} {165117} (\bibinfo {year}
  {2008})}\BibitemShut {NoStop}%
\bibitem [{\citenamefont {Go}\ \emph {et~al.}(2024)\citenamefont {Go},
  \citenamefont {Lee}, \citenamefont {Oppeneer}, \citenamefont {Bl\"ugel},\
  and\ \citenamefont {Mokrousov}}]{Go2024}%
  \BibitemOpen
  \bibfield  {author} {\bibinfo {author} {\bibfnamefont {D.}~\bibnamefont
  {Go}}, \bibinfo {author} {\bibfnamefont {H.-W.}\ \bibnamefont {Lee}},
  \bibinfo {author} {\bibfnamefont {P.~M.}\ \bibnamefont {Oppeneer}}, \bibinfo
  {author} {\bibfnamefont {S.}~\bibnamefont {Bl\"ugel}}, \ and\ \bibinfo
  {author} {\bibfnamefont {Y.}~\bibnamefont {Mokrousov}},\ }\bibfield  {title}
  {\enquote {\bibinfo {title} {First-principles calculation of orbital hall
  effect by wannier interpolation: Role of orbital dependence of the anomalous
  position},}\ }\href {\doibase 10.1103/PhysRevB.109.174435} {\bibfield
  {journal} {\bibinfo  {journal} {Phys. Rev. B}\ }\textbf {\bibinfo {volume}
  {109}},\ \bibinfo {pages} {174435} (\bibinfo {year} {2024})}\BibitemShut
  {NoStop}%
\bibitem [{\citenamefont {Sagasta}\ \emph {et~al.}(2016)\citenamefont
  {Sagasta}, \citenamefont {Omori}, \citenamefont {Isasa}, \citenamefont
  {Gradhand}, \citenamefont {Hueso}, \citenamefont {Niimi}, \citenamefont
  {Otani},\ and\ \citenamefont {Casanova}}]{Sagasta2016}%
  \BibitemOpen
  \bibfield  {author} {\bibinfo {author} {\bibfnamefont {E.}~\bibnamefont
  {Sagasta}}, \bibinfo {author} {\bibfnamefont {Y.}~\bibnamefont {Omori}},
  \bibinfo {author} {\bibfnamefont {M.}~\bibnamefont {Isasa}}, \bibinfo
  {author} {\bibfnamefont {M.}~\bibnamefont {Gradhand}}, \bibinfo {author}
  {\bibfnamefont {L.~E.}\ \bibnamefont {Hueso}}, \bibinfo {author}
  {\bibfnamefont {Y.}~\bibnamefont {Niimi}}, \bibinfo {author} {\bibfnamefont
  {Y.}~\bibnamefont {Otani}}, \ and\ \bibinfo {author} {\bibfnamefont
  {F.}~\bibnamefont {Casanova}},\ }\bibfield  {title} {\enquote {\bibinfo
  {title} {Tuning the spin hall effect of pt from the moderately dirty to the
  superclean regime},}\ }\href {\doibase 10.1103/PhysRevB.94.060412} {\bibfield
   {journal} {\bibinfo  {journal} {Physical Review B}\ }\textbf {\bibinfo
  {volume} {94}},\ \bibinfo {pages} {060412} (\bibinfo {year}
  {2016})}\BibitemShut {NoStop}%
\bibitem [{\citenamefont {Meinert}\ \emph {et~al.}(2020)\citenamefont
  {Meinert}, \citenamefont {Gliniors}, \citenamefont {Gueckstock},
  \citenamefont {Seifert}, \citenamefont {Liensberger}, \citenamefont {Weiler},
  \citenamefont {Wimmer}, \citenamefont {Ebert},\ and\ \citenamefont
  {Kampfrath}}]{Meinert2020}%
  \BibitemOpen
  \bibfield  {author} {\bibinfo {author} {\bibfnamefont {M.}~\bibnamefont
  {Meinert}}, \bibinfo {author} {\bibfnamefont {B.}~\bibnamefont {Gliniors}},
  \bibinfo {author} {\bibfnamefont {O.}~\bibnamefont {Gueckstock}}, \bibinfo
  {author} {\bibfnamefont {T.~S.}\ \bibnamefont {Seifert}}, \bibinfo {author}
  {\bibfnamefont {L.}~\bibnamefont {Liensberger}}, \bibinfo {author}
  {\bibfnamefont {M.}~\bibnamefont {Weiler}}, \bibinfo {author} {\bibfnamefont
  {S.}~\bibnamefont {Wimmer}}, \bibinfo {author} {\bibfnamefont
  {H.}~\bibnamefont {Ebert}}, \ and\ \bibinfo {author} {\bibfnamefont
  {T.}~\bibnamefont {Kampfrath}},\ }\bibfield  {title} {\enquote {\bibinfo
  {title} {High-throughput techniques for measuring the spin hall effect},}\
  }\href {\doibase 10.1103/PhysRevApplied.14.064011} {\bibfield  {journal}
  {\bibinfo  {journal} {Physical Review Applied}\ }\textbf {\bibinfo {volume}
  {14}},\ \bibinfo {pages} {064011} (\bibinfo {year} {2020})}\BibitemShut
  {NoStop}%
\bibitem [{\citenamefont {McHugh}, \citenamefont {Gradhand},\ and\
  \citenamefont {Stewart}(2024)}]{McHugh2024}%
  \BibitemOpen
  \bibfield  {author} {\bibinfo {author} {\bibfnamefont {O.~L.}\ \bibnamefont
  {McHugh}}, \bibinfo {author} {\bibfnamefont {M.}~\bibnamefont {Gradhand}}, \
  and\ \bibinfo {author} {\bibfnamefont {D.~A.}\ \bibnamefont {Stewart}},\
  }\bibfield  {title} {\enquote {\bibinfo {title} {Optimizing the spin hall
  effect in pt-based binary alloys},}\ }\href {\doibase
  10.1103/PhysRevMaterials.8.015003} {\bibfield  {journal} {\bibinfo  {journal}
  {Physical Review Materials}\ }\textbf {\bibinfo {volume} {8}} (\bibinfo
  {year} {2024}),\ 10.1103/PhysRevMaterials.8.015003}\BibitemShut {NoStop}%
\bibitem [{\citenamefont {Lau}, \citenamefont {Seki},\ and\ \citenamefont
  {Takanashi}(2021)}]{Lau2021}%
  \BibitemOpen
  \bibfield  {author} {\bibinfo {author} {\bibfnamefont {Y.-C.}\ \bibnamefont
  {Lau}}, \bibinfo {author} {\bibfnamefont {T.}~\bibnamefont {Seki}}, \ and\
  \bibinfo {author} {\bibfnamefont {K.}~\bibnamefont {Takanashi}},\ }\bibfield
  {title} {\enquote {\bibinfo {title} {Highly fcc-textured pt–al alloy films
  grown on mgo(001) showing enhanced spin hall efficiency},}\ }\href {\doibase
  10.1063/5.0052544} {\bibfield  {journal} {\bibinfo  {journal} {APL
  Materials}\ }\textbf {\bibinfo {volume} {9}},\ \bibinfo {pages} {081113}
  (\bibinfo {year} {2021})},\ \Eprint
  {http://arxiv.org/abs/https://pubs.aip.org/aip/apm/article-pdf/doi/10.1063/5.0052544/14566062/081113\_1\_online.pdf}
  {https://pubs.aip.org/aip/apm/article-pdf/doi/10.1063/5.0052544/14566062/081113\_1\_online.pdf}
  \BibitemShut {NoStop}%
\bibitem [{\citenamefont {McAlister}\ and\ \citenamefont
  {Kahan}(1986)}]{McAlister1986}%
  \BibitemOpen
  \bibfield  {author} {\bibinfo {author} {\bibfnamefont {A.~J.}\ \bibnamefont
  {McAlister}}\ and\ \bibinfo {author} {\bibfnamefont {D.~J.}\ \bibnamefont
  {Kahan}},\ }\bibfield  {title} {\enquote {\bibinfo {title} {The al-pt
  (aluminum-platinum) system},}\ }\href {\doibase 10.1007/BF02874982}
  {\bibfield  {journal} {\bibinfo  {journal} {Bulletin of Alloy Phase
  Diagrams}\ }\textbf {\bibinfo {volume} {7}},\ \bibinfo {pages} {47--51}
  (\bibinfo {year} {1986})}\BibitemShut {NoStop}%
\bibitem [{\citenamefont {Jain}\ \emph {et~al.}(2013)\citenamefont {Jain},
  \citenamefont {Ong}, \citenamefont {Hautier}, \citenamefont {Chen},
  \citenamefont {Richards}, \citenamefont {Dacek}, \citenamefont {Cholia},
  \citenamefont {Gunter}, \citenamefont {Skinner}, \citenamefont {Ceder},\ and\
  \citenamefont {Persson}}]{Jain2013}%
  \BibitemOpen
  \bibfield  {author} {\bibinfo {author} {\bibfnamefont {A.}~\bibnamefont
  {Jain}}, \bibinfo {author} {\bibfnamefont {S.~P.}\ \bibnamefont {Ong}},
  \bibinfo {author} {\bibfnamefont {G.}~\bibnamefont {Hautier}}, \bibinfo
  {author} {\bibfnamefont {W.}~\bibnamefont {Chen}}, \bibinfo {author}
  {\bibfnamefont {W.~D.}\ \bibnamefont {Richards}}, \bibinfo {author}
  {\bibfnamefont {S.}~\bibnamefont {Dacek}}, \bibinfo {author} {\bibfnamefont
  {S.}~\bibnamefont {Cholia}}, \bibinfo {author} {\bibfnamefont
  {D.}~\bibnamefont {Gunter}}, \bibinfo {author} {\bibfnamefont
  {D.}~\bibnamefont {Skinner}}, \bibinfo {author} {\bibfnamefont
  {G.}~\bibnamefont {Ceder}}, \ and\ \bibinfo {author} {\bibfnamefont {K.~A.}\
  \bibnamefont {Persson}},\ }\bibfield  {title} {\enquote {\bibinfo {title}
  {Commentary: The materials project: A materials genome approach to
  accelerating materials innovation},}\ }\href {\doibase 10.1063/1.4812323}
  {\bibfield  {journal} {\bibinfo  {journal} {APL Materials}\ }\textbf
  {\bibinfo {volume} {1}},\ \bibinfo {pages} {011002} (\bibinfo {year}
  {2013})},\ \Eprint
  {http://arxiv.org/abs/https://pubs.aip.org/aip/apm/article-pdf/doi/10.1063/1.4812323/13163869/011002\_1\_online.pdf}
  {https://pubs.aip.org/aip/apm/article-pdf/doi/10.1063/1.4812323/13163869/011002\_1\_online.pdf}
  \BibitemShut {NoStop}%
\bibitem [{\citenamefont {Ebert}, \citenamefont {Ködderitzsch},\ and\
  \citenamefont {Minár}(2011)}]{Ebert2011}%
  \BibitemOpen
  \bibfield  {author} {\bibinfo {author} {\bibfnamefont {H.}~\bibnamefont
  {Ebert}}, \bibinfo {author} {\bibfnamefont {D.}~\bibnamefont
  {Ködderitzsch}}, \ and\ \bibinfo {author} {\bibfnamefont {J.}~\bibnamefont
  {Minár}},\ }\bibfield  {title} {\enquote {\bibinfo {title} {Calculating
  condensed matter properties using the kkr-green's function method—recent
  developments and applications},}\ }\href {\doibase
  10.1088/0034-4885/74/9/096501} {\bibfield  {journal} {\bibinfo  {journal}
  {Reports on Progress in Physics}\ }\textbf {\bibinfo {volume} {74}},\
  \bibinfo {pages} {096501} (\bibinfo {year} {2011})}\BibitemShut {NoStop}%
\bibitem [{\citenamefont {Lowitzer}\ \emph {et~al.}(2011)\citenamefont
  {Lowitzer}, \citenamefont {Gradhand}, \citenamefont {Ködderitzsch},
  \citenamefont {Fedorov}, \citenamefont {Mertig},\ and\ \citenamefont
  {Ebert}}]{Lowitzer2011}%
  \BibitemOpen
  \bibfield  {author} {\bibinfo {author} {\bibfnamefont {S.}~\bibnamefont
  {Lowitzer}}, \bibinfo {author} {\bibfnamefont {M.}~\bibnamefont {Gradhand}},
  \bibinfo {author} {\bibfnamefont {D.}~\bibnamefont {Ködderitzsch}}, \bibinfo
  {author} {\bibfnamefont {D.~V.}\ \bibnamefont {Fedorov}}, \bibinfo {author}
  {\bibfnamefont {I.}~\bibnamefont {Mertig}}, \ and\ \bibinfo {author}
  {\bibfnamefont {H.}~\bibnamefont {Ebert}},\ }\bibfield  {title} {\enquote
  {\bibinfo {title} {Extrinsic and intrinsic contributions to the spin hall
  effect of alloys},}\ }\href {\doibase 10.1103/PhysRevLett.106.056601}
  {\bibfield  {journal} {\bibinfo  {journal} {Physical Review Letters}\
  }\textbf {\bibinfo {volume} {106}},\ \bibinfo {pages} {056601} (\bibinfo
  {year} {2011})}\BibitemShut {NoStop}%
\bibitem [{\citenamefont {Ebert}\ \emph {et~al.}(2015)\citenamefont {Ebert},
  \citenamefont {Mankovsky}, \citenamefont {Chadova}, \citenamefont {Polesya},
  \citenamefont {Minár},\ and\ \citenamefont {Ködderitzsch}}]{Ebert2015}%
  \BibitemOpen
  \bibfield  {author} {\bibinfo {author} {\bibfnamefont {H.}~\bibnamefont
  {Ebert}}, \bibinfo {author} {\bibfnamefont {S.}~\bibnamefont {Mankovsky}},
  \bibinfo {author} {\bibfnamefont {K.}~\bibnamefont {Chadova}}, \bibinfo
  {author} {\bibfnamefont {S.}~\bibnamefont {Polesya}}, \bibinfo {author}
  {\bibfnamefont {J.}~\bibnamefont {Minár}}, \ and\ \bibinfo {author}
  {\bibfnamefont {D.}~\bibnamefont {Ködderitzsch}},\ }\bibfield  {title}
  {\enquote {\bibinfo {title} {Calculating linear-response functions for finite
  temperatures on the basis of the alloy analogy model},}\ }\href {\doibase
  10.1103/PhysRevB.91.165132} {\bibfield  {journal} {\bibinfo  {journal}
  {Physical Review B}\ }\textbf {\bibinfo {volume} {91}},\ \bibinfo {pages}
  {165132} (\bibinfo {year} {2015})}\BibitemShut {NoStop}%
\bibitem [{\citenamefont {Fritz}\ \emph {et~al.}(2018)\citenamefont {Fritz},
  \citenamefont {Wimmer}, \citenamefont {Ebert},\ and\ \citenamefont
  {Meinert}}]{Fritz2018}%
  \BibitemOpen
  \bibfield  {author} {\bibinfo {author} {\bibfnamefont {K.}~\bibnamefont
  {Fritz}}, \bibinfo {author} {\bibfnamefont {S.}~\bibnamefont {Wimmer}},
  \bibinfo {author} {\bibfnamefont {H.}~\bibnamefont {Ebert}}, \ and\ \bibinfo
  {author} {\bibfnamefont {M.}~\bibnamefont {Meinert}},\ }\bibfield  {title}
  {\enquote {\bibinfo {title} {Large spin hall effect in an amorphous binary
  alloy},}\ }\href {\doibase 10.1103/PhysRevB.98.094433} {\bibfield  {journal}
  {\bibinfo  {journal} {Physical Review B}\ }\textbf {\bibinfo {volume} {98}},\
  \bibinfo {pages} {094433} (\bibinfo {year} {2018})}\BibitemShut {NoStop}%
\bibitem [{\citenamefont {Chadova}\ \emph {et~al.}(2015)\citenamefont
  {Chadova}, \citenamefont {Fedorov}, \citenamefont {Herschbach}, \citenamefont
  {Gradhand}, \citenamefont {Mertig}, \citenamefont {K\"odderitzsch},\ and\
  \citenamefont {Ebert}}]{Chadova2015}%
  \BibitemOpen
  \bibfield  {author} {\bibinfo {author} {\bibfnamefont {K.}~\bibnamefont
  {Chadova}}, \bibinfo {author} {\bibfnamefont {D.~V.}\ \bibnamefont
  {Fedorov}}, \bibinfo {author} {\bibfnamefont {C.}~\bibnamefont {Herschbach}},
  \bibinfo {author} {\bibfnamefont {M.}~\bibnamefont {Gradhand}}, \bibinfo
  {author} {\bibfnamefont {I.}~\bibnamefont {Mertig}}, \bibinfo {author}
  {\bibfnamefont {D.}~\bibnamefont {K\"odderitzsch}}, \ and\ \bibinfo {author}
  {\bibfnamefont {H.}~\bibnamefont {Ebert}},\ }\bibfield  {title} {\enquote
  {\bibinfo {title} {Separation of the individual contributions to the spin
  hall effect in dilute alloys within the first-principles kubo-st\ifmmode
  \check{r}\else \v{r}\fi{}eda approach},}\ }\href {\doibase
  10.1103/PhysRevB.92.045120} {\bibfield  {journal} {\bibinfo  {journal} {Phys.
  Rev. B}\ }\textbf {\bibinfo {volume} {92}},\ \bibinfo {pages} {045120}
  (\bibinfo {year} {2015})}\BibitemShut {NoStop}%
\bibitem [{\citenamefont {Dutta}\ \emph {et~al.}(2017)\citenamefont {Dutta},
  \citenamefont {Sankaran}, \citenamefont {Moors}, \citenamefont {Pourtois},
  \citenamefont {Elshocht}, \citenamefont {Bömmels}, \citenamefont
  {Vandervorst}, \citenamefont {Tőkei},\ and\ \citenamefont
  {Adelmann}}]{Dutta2017}%
  \BibitemOpen
  \bibfield  {author} {\bibinfo {author} {\bibfnamefont {S.}~\bibnamefont
  {Dutta}}, \bibinfo {author} {\bibfnamefont {K.}~\bibnamefont {Sankaran}},
  \bibinfo {author} {\bibfnamefont {K.}~\bibnamefont {Moors}}, \bibinfo
  {author} {\bibfnamefont {G.}~\bibnamefont {Pourtois}}, \bibinfo {author}
  {\bibfnamefont {S.~V.}\ \bibnamefont {Elshocht}}, \bibinfo {author}
  {\bibfnamefont {J.}~\bibnamefont {Bömmels}}, \bibinfo {author}
  {\bibfnamefont {W.}~\bibnamefont {Vandervorst}}, \bibinfo {author}
  {\bibfnamefont {Z.}~\bibnamefont {Tőkei}}, \ and\ \bibinfo {author}
  {\bibfnamefont {C.}~\bibnamefont {Adelmann}},\ }\bibfield  {title} {\enquote
  {\bibinfo {title} {Thickness dependence of the resistivity of platinum-group
  metal thin films},}\ }\href {\doibase 10.1063/1.4992089} {\bibfield
  {journal} {\bibinfo  {journal} {Journal of Applied Physics}\ }\textbf
  {\bibinfo {volume} {122}},\ \bibinfo {pages} {025107} (\bibinfo {year}
  {2017})}\BibitemShut {NoStop}%
\end{thebibliography}%

\end{document}